\documentclass[a4paper]{jpconf}
\usepackage{graphicx}
\usepackage{amssymb}

\def\slashchar#1{\setbox0=\hbox{$#1$}
   \dimen0=\wd0 \setbox1=\hbox{/} \dimen1=\wd1
   \ifdim\dimen0>\dimen1 \rlap{\hbox to \dimen0{\hfil/\hfil}} #1
   \else  \rlap{\hbox to \dimen1{\hfil$#1$\hfil}} / \fi}

\begin{document}
\title{Weak Strangeness and Eta Production}
\author{M. Rafi Alam, M. Sajjad Athar}
\address{Department of Physics, Aligarh Muslim University, Aligarh, India 202002}
\ead{rafi.alam.amu@gmail.com}

\author{Luis Alvarez-Ruso, I Ruiz Simo, M.J. Vicente Vacas}
\address{Departamento de F\'\i sica Te\'orica e IFIC, Centro Mixto
Universidad de Valencia-CSIC, Institutos de Investigaci\'on de
Paterna, Apartado 22085, E-46071 Valencia, Spain}

\author{S. K. Singh}
\address{H. N. B. Garhwal University, Srinagar - 246 174, India}

\begin{abstract}
We have studied strange particle production off nucleons through $\Delta S =0 $ and $|\Delta S| = 1$ channels, and specifically single
 kaon/antikaon, eta, associated particle production for neutrino/antineutrino induced processes as well as antineutrino induced single hyperon production processes. 
 We have developed a microscopical model based on the SU(3) chiral Lagrangians. The basic parameters of the model are $f_\pi$, the pion decay constant, 
Cabibbo angle, the proton and neutron magnetic moments and the  axial vector coupling constants for the baryons octet. For antikaon production we have also included
 $\Sigma^*(1385)$ resonance and for eta production $S_{11}(1535)$ and $S_{11}(1650)$ resonances are included.
\end{abstract}

\section{Introduction}
In the intermediate energy region of a few GeV, the cross sections for charged lepton production
are dominated by quasielastic and various inelastic processes involving pion production induced by charged 
and neutral currents in neutrino and antineutrino reactions from nuclear targets. 
The experimental observations of strange particles through weak interaction induced $\Delta S =0 $ and $|\Delta S| = 1 $ processes are quite limited. 
These are limited both by statistics as well as by the large systematic errors. Also theoretically there are only a few works available in literature. 
However, the availability of high intensity neutrino and antineutrino beams in present generation neutrino 
experiments has opened up the possibility of experimentally studying these processes with better statistics. 
In this work, we summarize our results of the total scattering cross sections for single kaon/antikaon, eta, associated particle production 
for neutrino/antineutrino induced processes as well as antineutrino induced single hyperon production. The details of the model are given in 
Refs.~\cite{RafiAlam:2010kf,Alam:2012zz,Alam:2013,Alam:2013woa,Singh:2006xp}. 
The calculations are performed using a microscopical model based on the 
chiral perturbation theory($\chi PT$) at the level of baryon and meson octet~\cite{RafiAlam:2010kf} for the background terms and for $K^-/{\bar K^0}$ production resonant
mechanism is introduced by the inclusion of the 
decuplet baryons~\cite{Alam:2012zz}. 
For $|\Delta S| =1 $ hyperon production cross section induced by antineutrinos we followed the prescription 
of Ref.~\cite{Singh:2006xp}. 

\section{Results and Discussions}
\begin{figure}[t]
\begin{center}
\includegraphics[height=5.5cm, width=7.5cm]{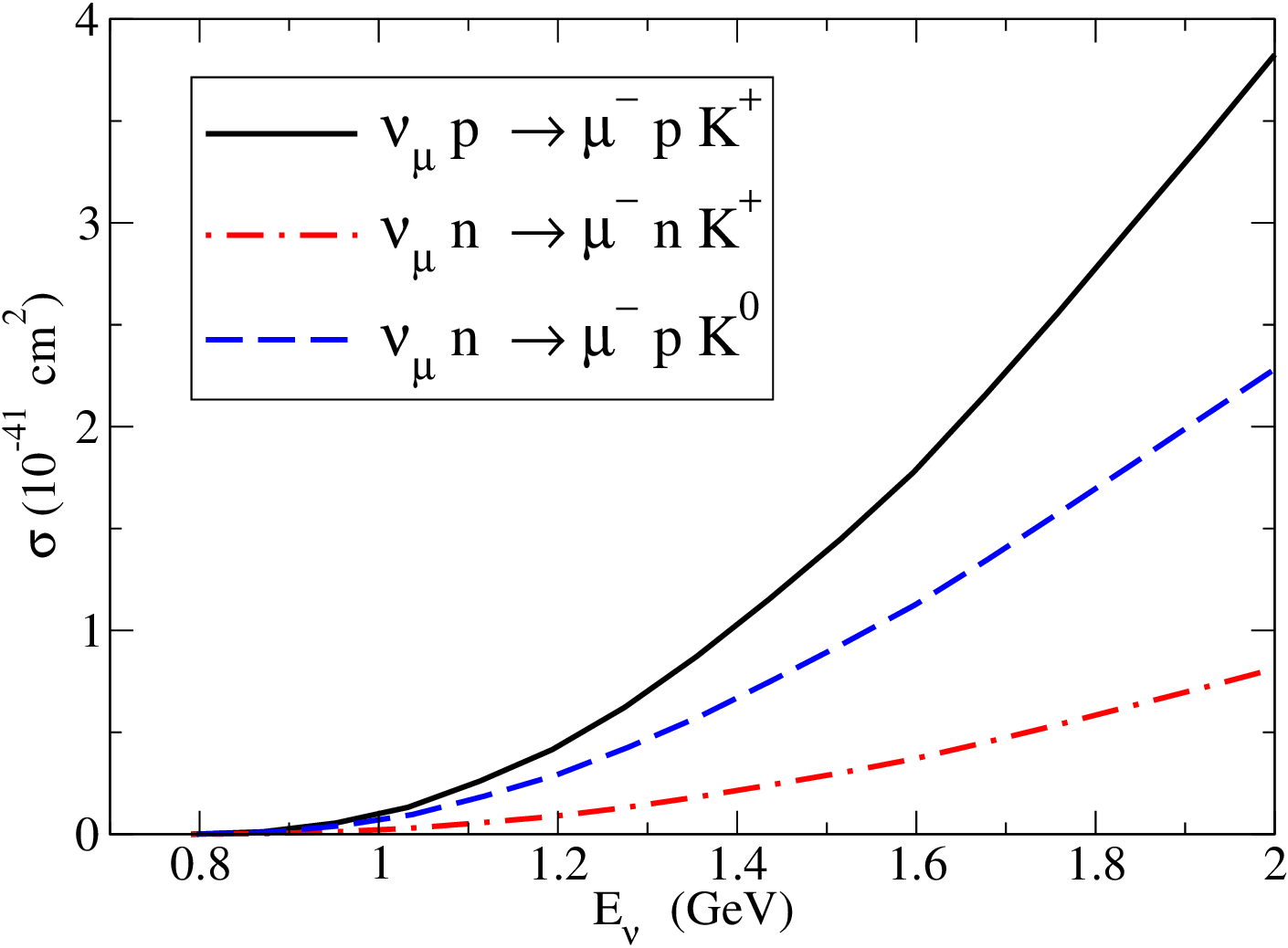}
\includegraphics[height=5.5cm, width=7.5cm]{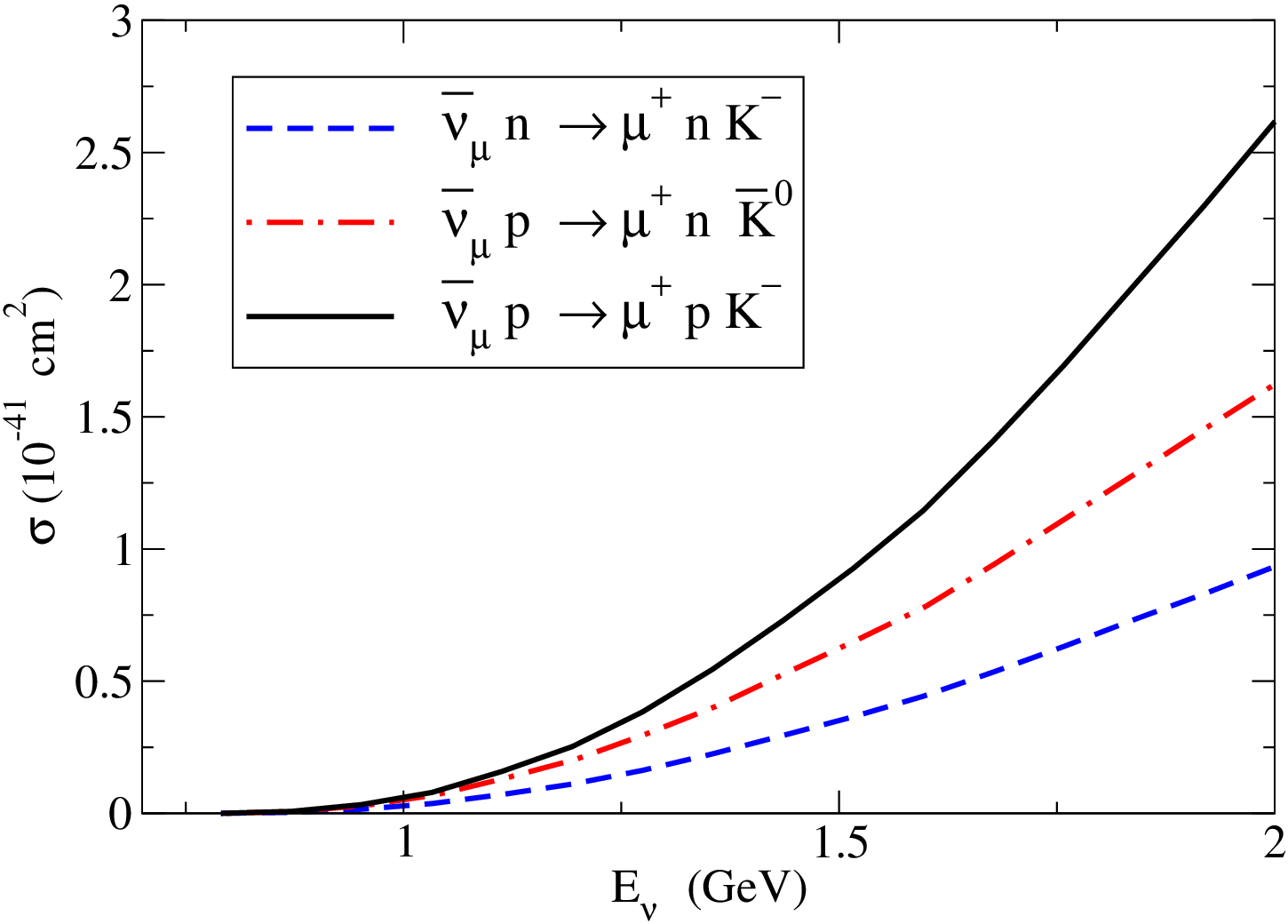}
\caption{$\sigma$ vs $E_{\nu_\mu/ \bar \nu_\mu}$ for $|\Delta S|=1$, K(left panel) and $\bar K$(right panel) production.}
\label{fig:total_single_xsec}
\end{center}
\end{figure}
\begin{figure}[t]
\begin{center}
\includegraphics[height=5.5cm, width=7.5cm]{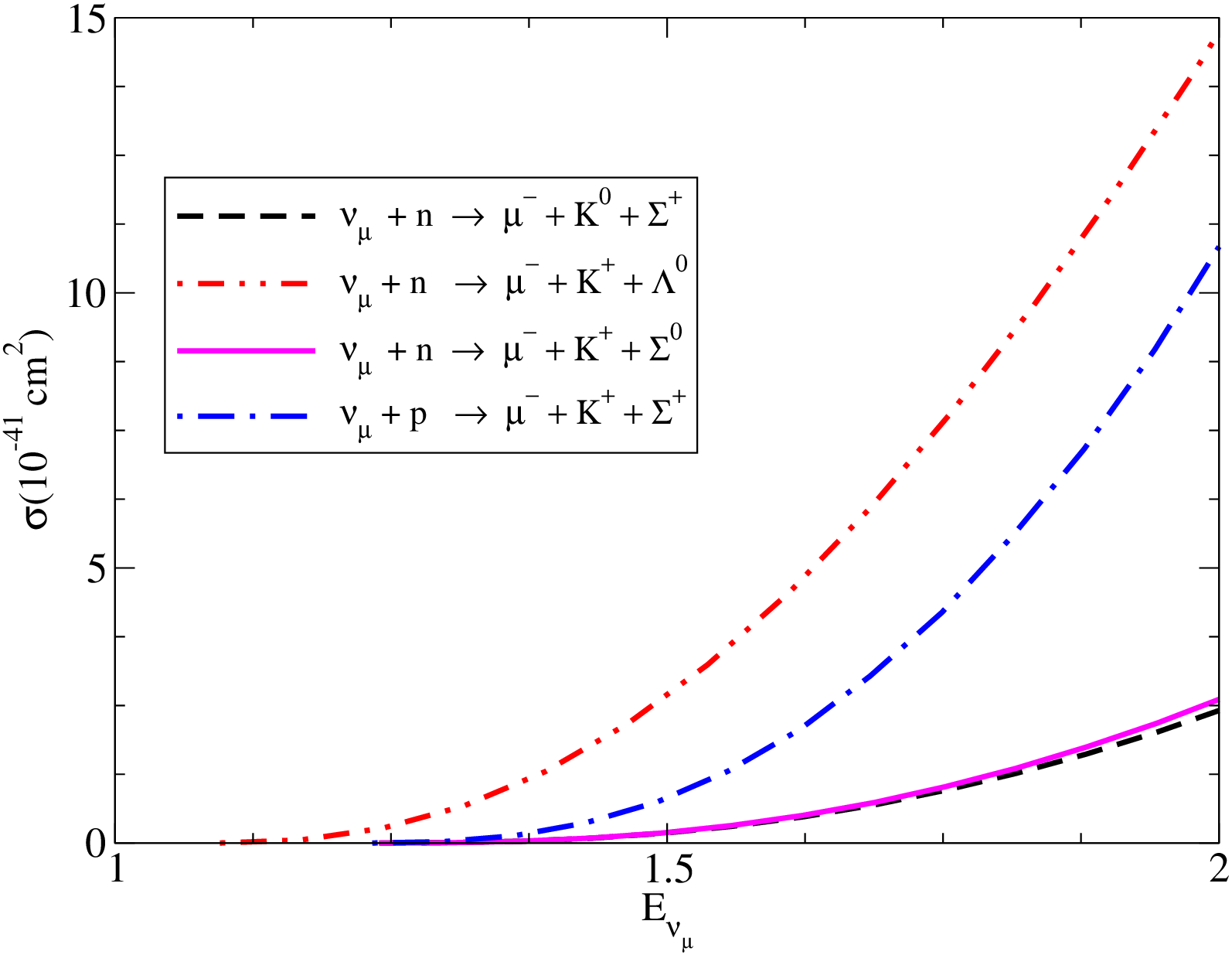}
\includegraphics[height=5.5cm, width=7.5cm]{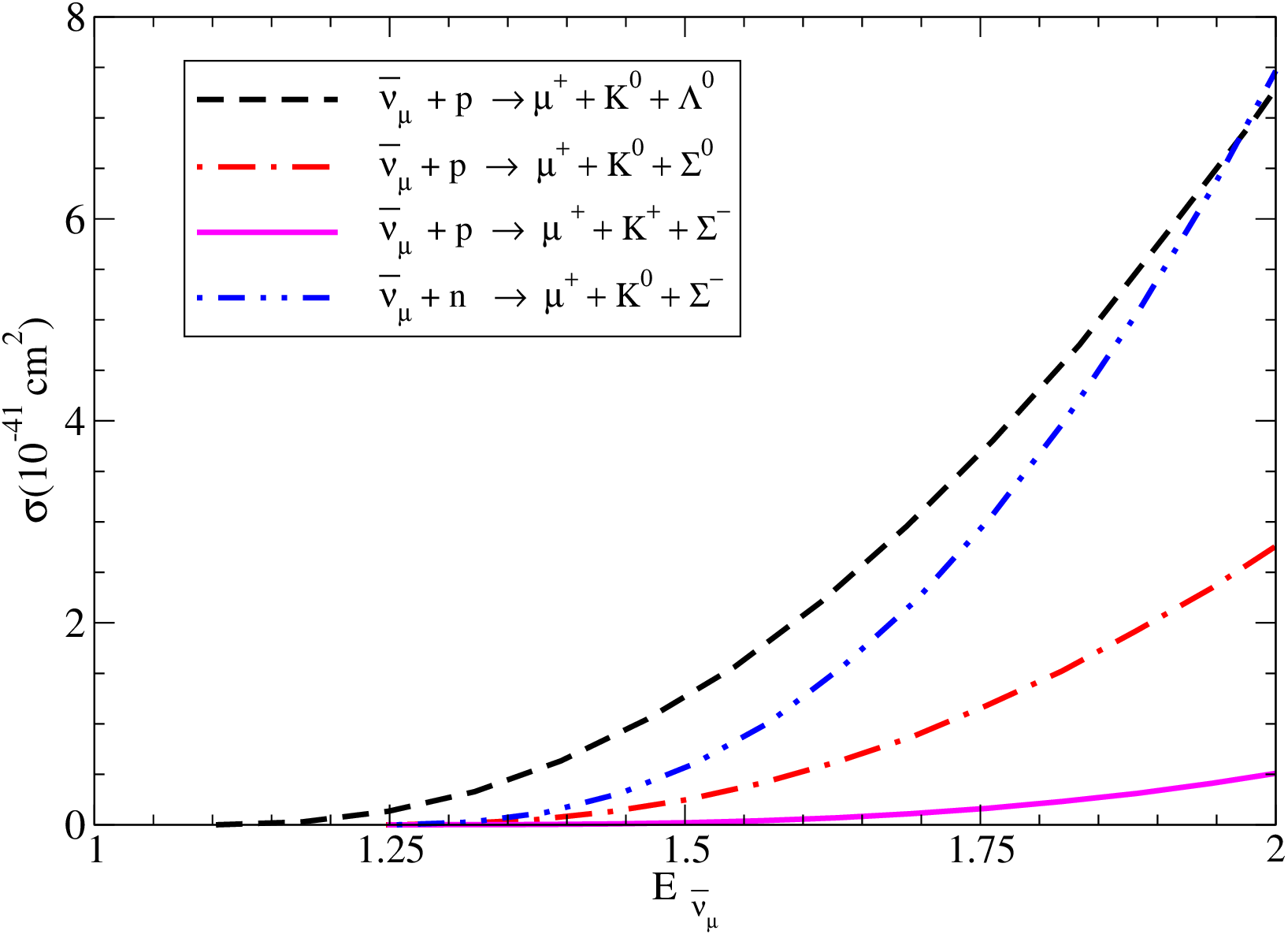}
\caption{$\sigma$ vs $E_{\nu_\mu/ \bar \nu_\mu}$ for $\Delta S=0$ process induced by $\nu$(left panel) and 
$\bar\nu$(right panel).}
\label{fig:total_ass_xsec}
\end{center}
\end{figure}
 Here we are discussing some of the results for $\Delta S =0 $ and $|\Delta S| = 1$ processes, 
which would be quite useful in the analysis of neutrino oscillation physics, in estimation of the background in proton decay searches besides their own intrinsic importance.
For single kaon production ($|\Delta S| = 1$), we have considered the following reactions:
\begin{small}
\begin{eqnarray}\label{kaon}
\nu_\mu + p \rightarrow  \mu^- + K^+ + p ~~~~~~& \bar \nu_\mu + p \rightarrow  \mu^+ + K^- + p \nonumber \\
\nu_\mu + n \rightarrow \mu^- + K^0 + p ~~~~~~& \bar \nu_\mu + p \rightarrow  \mu^+ +\bar K^0 + n \nonumber \\ 
\nu_\mu + n \rightarrow \mu^- + K^+ + n ~~~~~~& \bar \nu_\mu + n \rightarrow  \mu^+ + K^- + n \; .
\end{eqnarray}
\end{small}
In Fig.~\ref{fig:total_single_xsec}, we have presented the results for $\nu(\bar\nu)$ induced K($\bar K$) production cross sections. 
 The kaon production gets contribution from contact term, kaon pole term, u-channel diagram and pion/eta in flight term~\cite{RafiAlam:2010kf}. 
 For the antikaon production besides the background terms, we have also taken the contribution from lowest lying $\Sigma^*(1385)$ resonance~\cite{Alam:2012zz}.

We find that the contact term is dominant, followed by the u-channel diagram with a $\Lambda$ intermediate state and the $\pi$ exchange term. 
The kaon pole contributions are negligible. We have used a global dipole form factor with a mass of 1 GeV and multiplied it with the hadronic current. 
We observe similar features in the results of antikaon production cross section as in the case of kaon production, 
and the contribution from $\Sigma^*(1385)$ resonance is very small.

In Fig.~\ref{fig:total_ass_xsec}, we have presented the results for neutrino and antineutrino induced associated kaon production cross section.
We have considered the following reactions for $\nu(\bar\nu)$ charged current induced associated 
particle production process:
\begin{small}
\begin{eqnarray}\label{associated}
\nu_\mu  + n \rightarrow \mu^-  + K^+  + \Lambda~~~~~~&\bar \nu_\mu  + p \rightarrow \mu^+ +  K^0  + 
\Lambda~~   \nonumber\\  
\nu_\mu  + p \rightarrow \mu^-  + K^+   + \Sigma^+ ~~~~~~& \bar \nu_\mu  + p \rightarrow \mu^+  + K^0  + 
\Sigma^0~ \nonumber\\ 
\nu_\mu  + n \rightarrow \mu^-  + K^+  + \Sigma^0~  ~~~~~~& \bar \nu_\mu  + p \rightarrow \mu^+  + 
K^+  + \Sigma^-  \nonumber\\  
\nu_\mu  + n \rightarrow \mu^-  + K^0  + \Sigma^+  ~~~~~~& \bar \nu_\mu  + n \rightarrow \mu^+  + K^0 
 + \Sigma^-   
\end{eqnarray}
\end{small}
The various form factors that appear in the vector and axial vector currents
have been determined using the prescription given in
Refs.~\cite{Singh:2006xp,Cabibbo:2003cu}. 
In this case also we find that the contact term has the largest contribution to
the cross section followed by s-channel and kaon pole terms. We find that the
cross section for the reaction channel with a $\Lambda$ in 
the final state are in general larger than that for the reactions where $\Sigma$
is in the 
final state. This can be understood by looking at the relative strength of the
coupling, for example, the ratio of couplings squared for the vertices 
$NK^0\Lambda$ to $NK^0\Sigma^0$ is $\frac{g^2_{NK\Lambda}}{g^2_{NK\Sigma}}
\simeq 14$. Furthermore, the cross section for the $\Lambda$ production is
favored by
the available phase space due to its small mass relative to $\Sigma$ baryons.
However, for the antineutrino induced process, 
the mechanism where $\Lambda$ is in the final state is not dominating. We
find that the neutrino induced reactions
are the dominant source of $K^+$ production whereas antineutrino induced reactions
favor $K^0$ production.

\begin{figure}[tbh]
\begin{center}
\includegraphics[height=5.5cm, width=7.5cm]{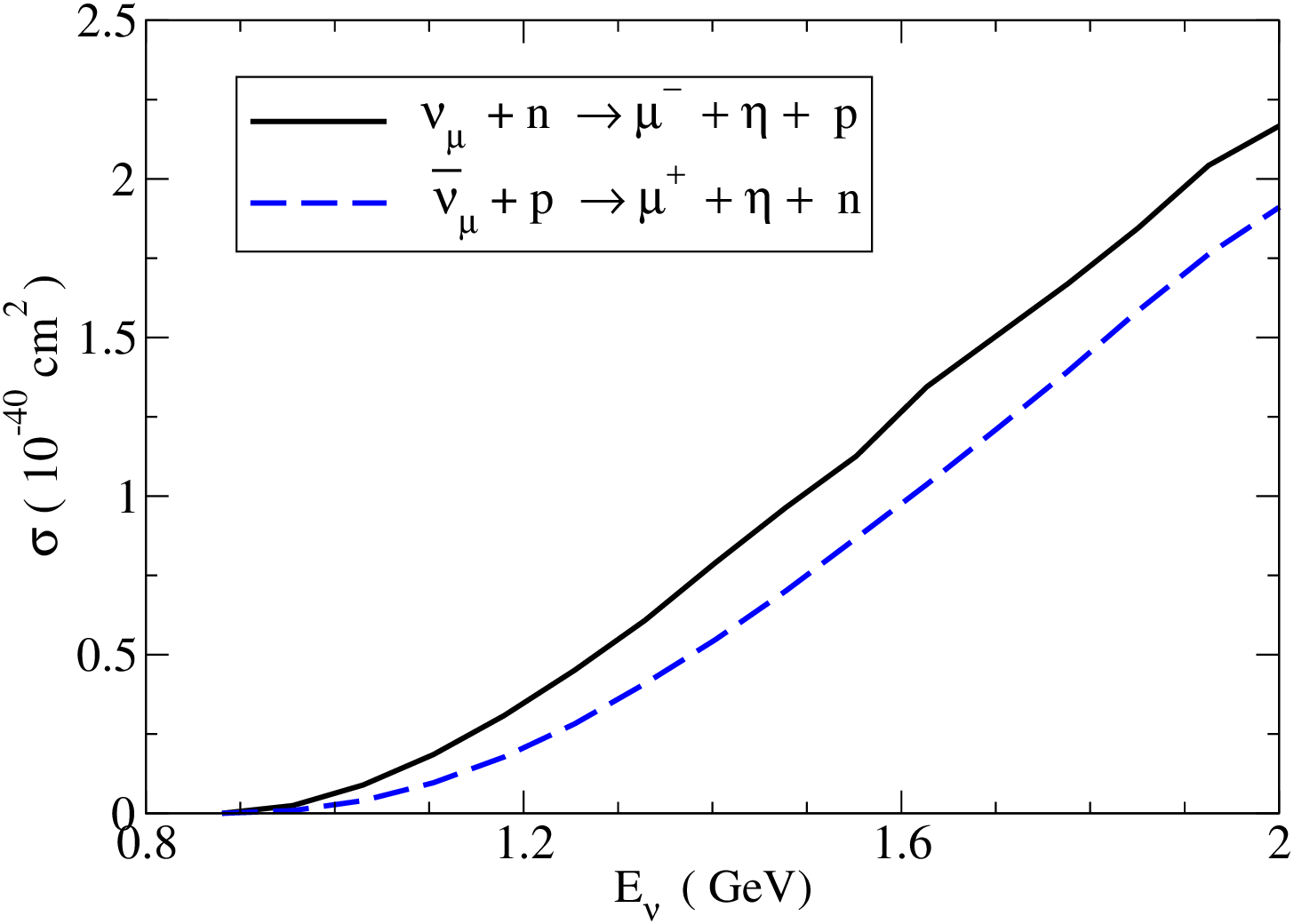}
\includegraphics[height=5.5cm, width=7.5cm]{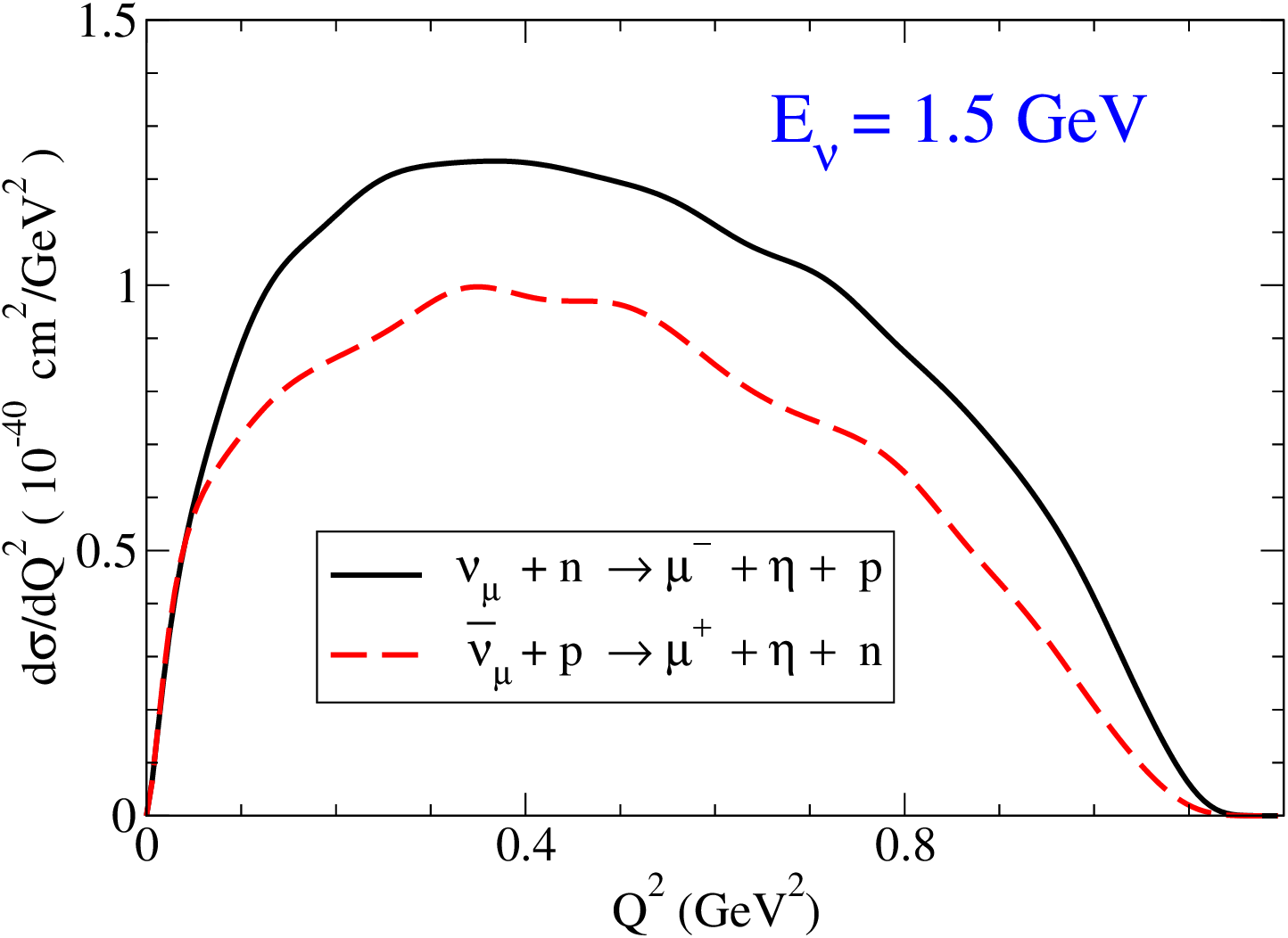}
\caption{$\sigma$(left panel) and the $Q^2$ distribution(right panel) for $\nu_\mu / \bar\nu_\mu $ induced $\eta$ production off the nucleon.}
\label{fig:eta_xsec}
\end{center}
\end{figure}

In Fig.~\ref{fig:eta_xsec}, we have presented the results for eta production cross section and $Q^2$ distribution, for the following processes:
\begin{small}
\begin{eqnarray}\label{eq:eta_weak_process}
\nu_\mu  + n \rightarrow \mu^- + \eta + p  ~~~~~~~~~~~~\bar \nu_\mu + p \rightarrow \mu^+ + \eta  + n
\end{eqnarray}
\end{small}
In these processes from symmetry only s- and u-channel nucleon pole terms contribute. Besides Born-terms, we have also considered 
$S_{11}$(1535) and $S_{11}$(1650) resonances. For the resonant mechanism we parameterized the vector part of the from factors using the 
helicity amplitudes~\cite{Alam:Prep}. We derived Goldberger-Treiman relation for the axial couplings and assumed a dipole 
form for $Q^2$ dependence for the axial form factors.
We find the dominance of $S_{11}$(1535) resonance followed by nucleon pole terms. $Q^2$ distribution in the case of eta production is almost 
flat in nature.  

At low energies, quasi-elastic production 
of hyperons induced by antineutrinos is possibile
\begin{small}
\begin{eqnarray}\label{reaction1}
\bar \nu_l + p \rightarrow l^+ + \Lambda ~~~~~~~~\bar \nu_l + p \rightarrow l^+ + \Sigma^0 ~~~~~~~~ \bar \nu_l + n \rightarrow l^+ + \Sigma^- 
\end{eqnarray}
\end{small}
\begin{figure}
\begin{center}
\includegraphics[height=6.5cm,width=15cm]{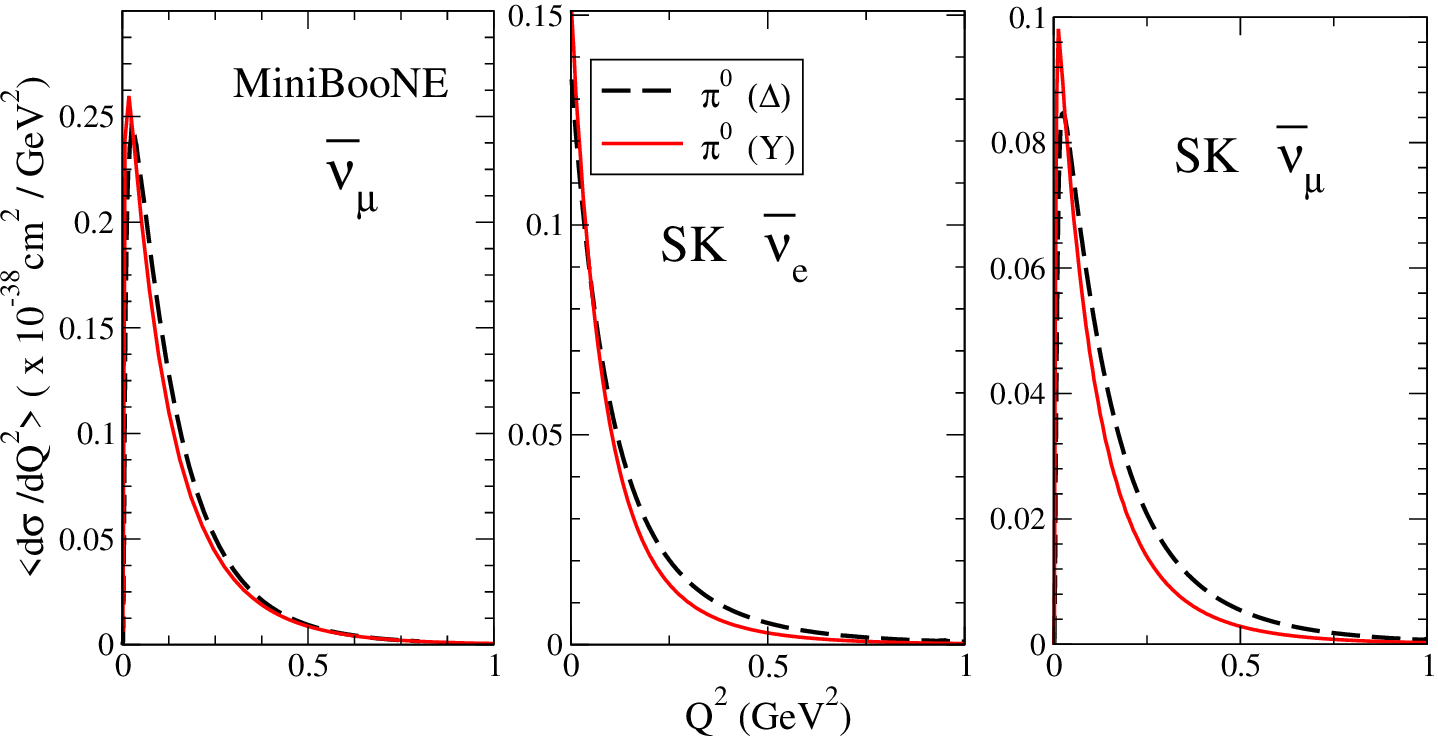}
\caption{$Q^2$ distributions for $\bar\nu_\mu$ induced reaction in $^{12}C$ averaged over the MiniBooNE flux,
and for $\bar\nu_{e,\mu}$ induced reaction in $^{16}O$ averaged over the atmospheric antineutrino flux at Super-Kamiokande 
are shown with nuclear medium and FSI effects for  $\pi^0$ production. The $\pi^0$ production 
from hyperon excitations have been scaled by a factor of 1.33.}
\label{fig:hyp_xsec}
\end{center}
\end{figure}
At the energies of MiniBooNE~\cite{AguilarArevalo:2008qa} and
atmospheric~\cite{Ashie:2005ik} $\bar\nu$ experiments, mainly pion production is
considered through the
 $\Delta S=0$ resonant mechanism with the dominance of $\Delta$ in the
intermediate state. In the case of antineutrino reactions in 
nucleon and nuclei there is an additional contribution to the pion production
from the quasielastic
$|\Delta S|=1$ processes mentioned in Eq.~\ref{reaction1}, in which hyperons
like $\Lambda, \Sigma^{-,0}$ can be produced and decay subsequently to pions. 
These processes are generally Cabibbo suppressed as compared to $\Delta$
production process
but could be important in the low energy region of antineutrinos. We find that
at low energies 
due to threshold effects and phase space considerations these processes give a
significant contribution to the $\pi$ production~\cite{Alam:2013}. 
In Fig.\ref{fig:hyp_xsec}, we present the results of $Q^2$ distribution averaged
over the MiniBooNE~\cite{AguilarArevalo:2008qa} and
atmospheric~\cite{Honda:2006qj}
antineutrino spectra, where $\pi^0$ production is obtained in the $\Delta$
dominance model following Ref.~\cite{Athar:2007wd} as well as $\pi^0$'s
contribution 
from hyperons following Ref.~\cite{Singh:2006xp}. These results are obtained in
$^{12}C$ for MiniBooNE and in $^{16}O$ for atmospheric antineutrinos. The effect of nuclear medium like Fermi motion 
and Pauli blocking is negligible. Furthermore, we find that when the contribution for the pions coming from all the hyperons
mentioned in Eq.\ref{reaction1} 
 is taken together there is no net change in the overall pion production due to final state interaction(FSI) effects, 
 as it increases in the $\Lambda$ production and decreases in the $\Sigma$ production in the nuclear medium~\cite{Alam:2013}. 
  We observe that in the peak region of $Q^2$ distribution, the contribution of
$\pi^0$ from the hyperon excitations is almost 75$\%$ to the
 contribution of $\pi^0$ from the $\Delta$ excitation. Similar is the
observation for $\pi^-$ production~\cite{Alam:2013}. Thus we find that for
antineutrino experiments 
 in the energy region of 0.6-1.0 GeV, the pion contribution from hyperons are
significant.
 
One of the authors(MSA) is thankful to PURSE program of D.S.T., Govt. of India
and the Aligarh Muslim University for the financial support. This research was
supported by the Spanish Ministerio de Economía y Competitividad and
European FEDER funds under Contracts FIS2011-28853-C02-01, by Generalitat
Valenciana under Contract No.
PROMETEO/20090090 and by the EU HadronPhysics3 project, Grant Agreement No.
283286.

\end{document}